\documentclass{nature}
\usepackage{graphicx}

\title{The remnants of galaxy formation from a panoramic survey of the region around M31}

\author{Alan W. McConnachie$^{1}$, Michael J. Irwin$^{2}$, Rodrigo A. Ibata$^{3}$, John Dubinski$^{4}$, Lawrence M. Widrow$^{5}$, Nicolas F. Martin$^{6}$, Patrick C{\^o}te$^{1}$, Aaron L. Dotter$^{7}$, Julio F. Navarro$^{7}$, Annette M. N. Ferguson$^{8}$, Thomas H. Puzia$^{1}$, Geraint F. Lewis$^{9}$, Arif Babul$^{7}$, Pauline Barmby$^{10}$, Olivier Bienayme$^{3}$, Scott C. Chapman$^{2}$, Robert Cockcroft$^{11}$, Michelle L. M. Collins$^{2}$, Mark A. Fardal$^{12}$, William E. Harris$^{11}$, Avon Huxor$^{13}$, A. Dougal Mackey$^{8}$, Jorge Pe{\~n}arrubia$^{2}$, R. Michael Rich$^{14}$, Harvey B. Richer$^{15}$, Arnaud Siebert$^{3}$, Nial Tanvir$^{16}$, David Valls-Gabaud$^{17}$ \& Kimberly A. Venn$^{7}$}

\begin{document}

\maketitle

\begin{affiliations}
 \item NRC Herzberg Institute of Astrophysics, 5071 West Saanich Road, Victoria, British Columbia, Canada V9E 2E7. 
\item Institute of Astronomy, University of Cambridge, Madingley Road, Cambridge CB3 0HA, UK. 
\item Observatoire de Strasbourg, 11, rue de l'Universit{\'e}, F-67000 Strasbourg, France.  
\item Department of Astronomy \& Astrophysics, University of Toronto, 50 St. George Street, Toronto, Ontario, Canada M5S 3H4. 
\item Department of Physics, Engineering Physics, and Astronomy Queen's University, Kingston, Ontario, Canada K7L 3N6. 
\item Max-Planck-Institut f{\"u}r Astronomie, K{\"o}nigstuhl 17, D-69117 Heidelberg, Germany. 
\item Department of Physics and Astronomy, University of Victoria, 3800 Finnerty Road, Victoria, British
Columbia, Canada V8P 5C2. 
\item Institute for Astronomy, University of Edinburgh, Royal Observatory, Blackford Hill, Edinburgh EH9 3HJ, UK. 
\item Sydney Institute for Astronomy, School of
Physics, University of Sydney, NSW2006, Australia. 
\item Department of Physics and Astronomy, University of Western Ontario, 1151 Richmond Street, London, Ontario, Canada N6A
3K7. 
\item Department of Physics and Astronomy,McMaster University, Hamilton, Ontario, Canada L8S 4M1. 
\item University of Massachusetts, Department of Astronomy, LGRT 619-E, 710
N. Pleasant Street, Amherst, Massachusetts 01003-9305, USA. 
\item Department of Physics (Astrophysics Group), H. H. Wills Physics Laboratory, Tyndall Avenue, Bristol BS8 1TL, UK.
\item Department of Physics and Astronomy, University of California, Los Angeles, PAB, 430 Portola Plaza, Los Angeles, California 90095-1547, USA. 
\item Department of Physics and
Astronomy, 6224 Agricultural Road, University of British Columbia, Vancouver, British Columbia, Canada V6T 1Z1. 
\item Department of Physics and Astronomy, University of Leicester,
Leicester LE1 7RH, UK. 
\item Laboratoire Galaxies et {'E}toiles, Physique et Instrumentation, CNRS UMR 8111, Observatoire de Paris, 5 Place Jules Janssen, 92195 Meudon, France.
\end{affiliations}

\newpage

\begin{abstract}
In hierarchical cosmological models$^{1}$, galaxies
grow in mass through the continual accretion of smaller ones. The
tidal disruption of these systems is expected to result in loosely
bound stars surrounding the galaxy, at distances that reach $10 - 100$
times the radius of the central
disk$^{2,3}$. The number, luminosity and
morphology of the relics of this process provide significant clues to
galaxy formation history$^{4}$, but obtaining a comprehensive survey of
these components is difficult because of their intrinsic faintness and
vast extent. Here we report a panoramic survey of the Andromeda galaxy
(M31). We detect stars and coherent structures that are almost
certainly remnants of dwarf galaxies destroyed by the tidal field of
M31. An improved census of their surviving counterparts implies that
three-quarters of M31's satellites brighter than $M_V < -6$ await
discovery. The brightest companion, Triangulum (M33), is surrounded by
a stellar structure that provides persuasive evidence for a recent
encounter with M31. This panorama of galaxy structure directly
confirms the basic tenets of the hierarchical galaxy formation model
and reveals the shared history of M31 and M33 in the unceasing
build-up of galaxies.
\end{abstract}

Precise measurements of stars and star clusters within the Milky Way
have contributed significantly to the development of a cosmological
model for galaxy formation$^{5,6}$. The discovery that the Sagittarius
dwarf galaxy was being cannibalized by the Milky Way$^{7}$ brought into
sharp focus the role of satellite accretion in the build-up of a
galaxy's mass. Models now propose that galaxies form within
dark-matter haloes that grow through the continual accretion and
merger of smaller sub-haloes. The predicted number of sub-haloes is at
least a few orders of magnitude more than the number of dwarf galaxies
observed as satellites around the Milky Way$^{8,9}$. If we assume that the
underlying cosmology is correct, this implies that either significant
numbers of satellites remain undiscovered or only a fraction of
sub-haloes contain baryons (stars and gas). In either case, the
number, luminosity and spatial distributions of satellite galaxies are
important but poorly determined quantities whose values depend
strongly on the processes through which baryons are accreted and
retained by sub-haloes$^{10,11}$. Many of the luminous sub-haloes are
expected to be perturbed, and even shredded, by the tidal field of the
host galaxy, leaving behind stellar debris in the form of streams and
substructures within a diffuse stellar halo$^{2,3}$.

Systematic studies of the Milky Way's stellar halo, such as with the
Sloan Digital Sky Survey$^{12}$, have recently revealed a large number of
dwarf galaxies and tidal streams. However, our viewpoint from within
the Milky Way introduces selection effects and difficulties in
interpretation, making a homogeneous global study difficult. To this
end, we have initiated the ``Pan-Andromeda Archaeological Survey''
(PAndAS), a programme using the 1-square-degree field-of-view
MegaPrime/MegaCam camera on the 3.6\,m Canada-France-Hawaii Telescope
(CFHT). We are imaging the closest spiral galaxy, M31, and its less
massive companion M33. Once completed, the survey will cover more than
300 square degrees (more than 70,000\,kpc$^2$) and extend to a maximum
projected radius from M31's centre of $r_p < 150$\,kpc. It is the
largest contiguous imaging survey of a massive galaxy and spans the
stellar halo out to extremely large radii. PAndAS builds on earlier
Isaac Newton Telescope and CFHT photometric studies of this
galaxy$^{13-17}$. It surveys in the $g$ ($4,140 - 5,600\AA$) and $i$
($7,020 - 8,530\AA$) bands and resolves individual stars in M31 to
depths of $g = 26.5$, $i = 25.5$ at a signal-to-noise ratio of 10. The
programme started in August 2008 and will continue until January 2011.

Figure~1 shows the spatial density distribution of sources in our
extant PAndAS fields that are consistent with red giant branch (RGB)
stars in M31. Although the region surveyed so far covers $\sim 220$ square
degrees (nearly 100-fold the area of the classical optical disk of M31),
we find RGB stars everywhere across our survey. These stars trace the
low-luminosity structure of the galaxy and reveal the vast extent of
M31, challenging the commonly held impression of the size of this
typical bright galaxy (and, by extension, other galaxies of similar
luminosities). These stars are unlikely to have formed in situ at these
radii because it is improbable that the local density of gas was high
enough to promote star formation. Instead, it is expected that the
stars have been accreted from dwarf galaxies or proto-galactic fragments,
a conclusion that is consistent with the basic tenets of hierarchical
galaxy formation.

Our interpretation of halo stars as accreted relics is supported by
the presence of multiple, large, coherent substructures over our
survey area. Some of these features, such as the giant stellar
stream$^{13}$ (no. 5 in Fig. 1), were previously known. New structures
discovered in our survey include a radial overdensity along the
northwest minor axis extending nearly 100 kpc from M31 (no. 6), a
diffuse structure to the southwest coherent over an arc spanning $\sim
40$\,kpc at a distance of $\sim 100$\,kpc from M31 (no. 7), and an
apparent continuation of a previously known stream in the east$^{16}$,
$\sim 50$\,kpc from M31 that loops around to the north of the galaxy
(no. 4). The large scale of the new structures is striking, as is
their distance from the centre of M31. They are expected to maintain
coherence for at least a few gigayears in the inner halo ($r_p <
50$\,kpc), and at least a Hubble time in the outer halo ($r_p >
100$\,kpc)$^{18}$. The implication is that these structures are the remains
of previously accreted dwarf galaxies. None of the newly discovered
stellar structures clearly correlate with HI detections in the
environs of M31 (ref. 19). Although the lack of young stars tracing
sites of star formation implies that high HI column densities are not
expected, more detailed predictions on the HI content of these
substructures are not yet possible.

The surviving counterparts of the progenitors of these substructures -
the dwarf galaxies - are visible in Fig. 1 as concentrated, round,
overdensities, and the large area surveyed permits an improved census
of these objects. In Fig. 2a we plot the projected radial number
density distribution of galaxies around M31; the profile shows no sign
of declining within 150\,kpc. In Fig. 2b we plot the cumulative
luminosity distribution of all galaxies around M31. We find that it is
well described by the usual Schechter function$^{20}$ with a faint-end
slope of $\alpha= 0.98 \pm 0.07$, where the uncertainty represents the
standard deviation. These distributions imply that $6 \pm 4$
satellites more luminous than $M_V \approx -6$ still remain to be
discovered within 150\,kpc (10\,degrees) of M31. Extrapolating the
observed flat number-density profile outwards then suggests that M31
may have as many as $88 \pm 20$ such satellites out to $r_p <
300$\,kpc (roughly equal to the expected virial radius of M31's
dark-matter halo; see Supplementary Information), only about
one-quarter of which are currently known. The derived number is an
upper limit, because we expect the radial profile to decline beyond
the survey area. Thus, even accounting for observational
incompleteness, there is still an order of magnitude too few
satellites compared with the expected number of dark-matter haloes. If
some of these haloes do not contain stars, then a comprehensive
inventory of those that do may shed much light on the solution to the
``missing satellites problem''.

The brightest of M31's satellite companions, M33, is surrounded by a
previously unknown prominent stellar structure (no. 1 in Fig. 1).
This feature has an extension stretching $\sim 2$\,degrees ($\sim
30$\,kpc projected) to the northwest towards M31, nearly three times
farther out than the classical disk of M33. A second extension is also
visible in the south.  One possible origin of this structure is that
it is the accreted remains of a dwarf satellite, similar to the
structures observed around M31.  However, a long-standing puzzle about
M33 is the existence of an extremely warped HI disk, with no apparent
counterpart in the ``pristine'' stellar disk$^{21,22}$. What process could
distort the HI disk but leave the stellar disk unaffected? The newly
discovered feature has a similar orientation to the HI warp,
suggesting that we have discovered its optical
counterpart. Furthermore, the general northwest - southeast symmetry
of the distortion, and the broad alignment of the northwest extension
with the measured transverse velocity of M33 (ref. 23), are evidence
of a tidal disturbance excited as this galaxy orbits around M31.

We test this hypothesis by considering the M33 - M31 orbit. If the
orbit of M33 carries it close to M31, the tidal field of the latter
will excite the M33 disk and eject stars, but the orbit cannot be so
close that M33 is severely distorted or disrupted$^{24}$. We use the
techniques described in the Supplementary Information to search for
M33 orbits that are consistent with the known constraints and have
reasonably close encounters with M31. The search reveals that the
smallest possible pericentre distance is $\sim 40$\,kpc, with close
passage occurring a few billion years ago. We select several
representative orbits and carry out n-body simulations$^{25}$ of the
encounter.

Figure 3 and Supplementary Movie 1 show results from one plausible
interaction model. This orbit reproduces with good accuracy the
observed distances$^{26}$, angular positions and radial velocities of M31
and M33, as well as the proper motion of M33 (ref. 23). The
interaction excites tidal tails in the M33 disk that wind up to form
an extended distribution of stars with dimensions similar to those of
the debris observed in the PAndAS data. In addition, the debris is
warped away from the disk in the same sense as the known gaseous
warp$^{21,22}$, lending further support to the interaction hypothesis.

Finally, we note that the encounter between M31 and M33 leaves an
imprint on the larger system, exciting a mild warp and disturbances in
the M31 disk at large radii. These phenomena are consistent with some
of the unusual features observed in the outer regions of M31
(ref. 14), particularly with the discovery of younger stellar
populations in these fields$^{27}$ and with the measured rotational
signature shared by many of the substructures$^{28}$. It is plausible that
these stars were originally formed in the thin disk and excited to
their present locations by a galactic interaction, perhaps with M33 at
an earlier phase in its orbit. The unrivalled panorama of galaxy
structure presented here for M31 therefore reveals the continuing role
of accretion and interactions in shaping its properties, and is a
startling visual demonstration of the truly vast scale of galaxies.

\begin{figure}
  \includegraphics[angle=0, width=17.cm]{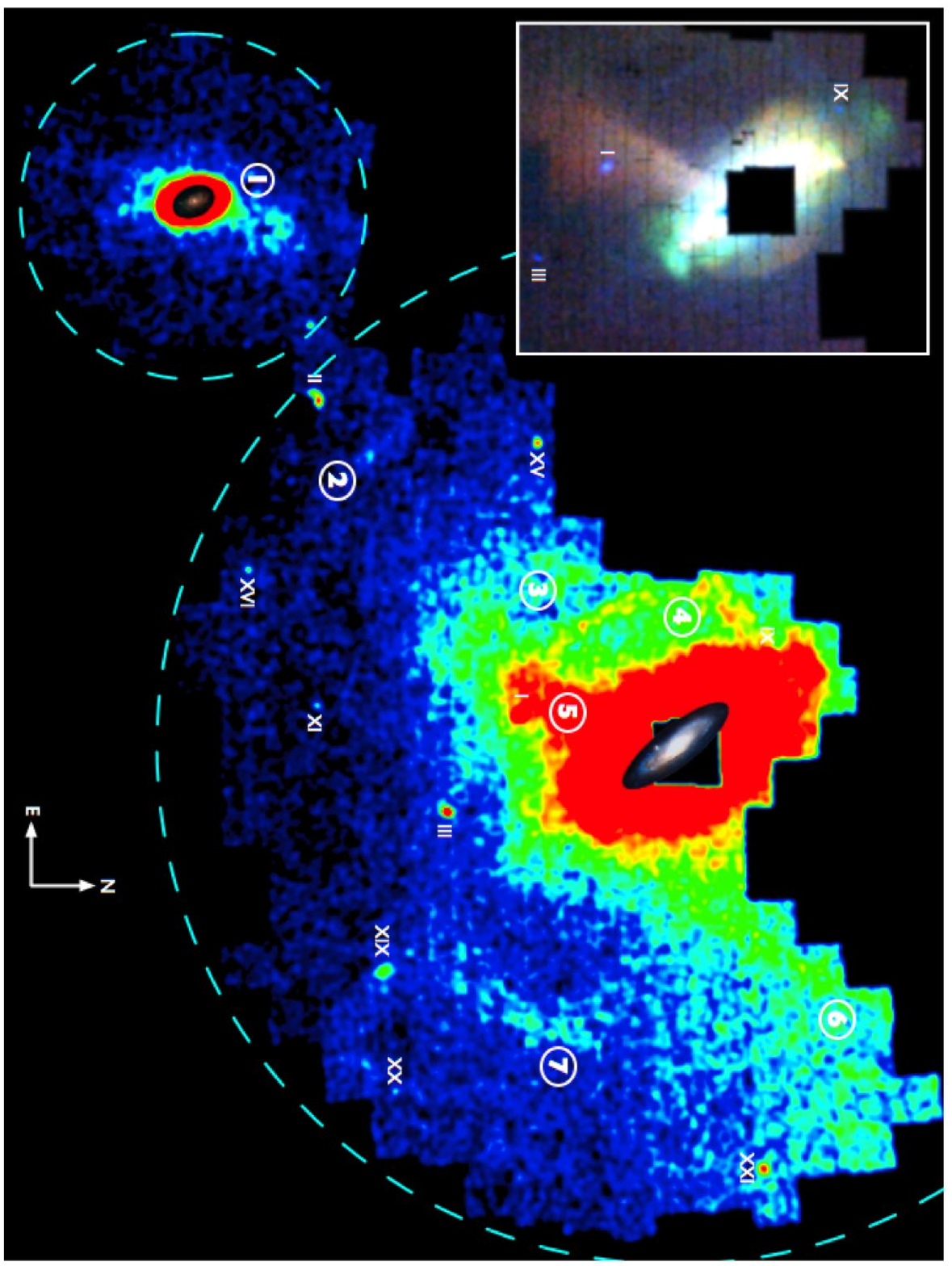}
\caption{{\bf Stellar density map of Andromeda -
    Triangulum.}  A tangent-plane projection of the density
  distribution of stellar sources in our extant PAndAS imaging is
  shown, with colours and magnitudes consistent with RGB stars at
  the distance of M31. The inset shows the central parts of our
  survey at higher resolution. Dashed circles represent the
  maximum projected radii of 150 and 50 kpc from M31 and M33,
  respectively. Scale images of the disks of M31 and M33 are
  overlaid. Visible dwarf satellites are indicated with roman
  numerals.  Numbers in circles indicate the largest and most
  obvious substructures detected in a visual inspection of the
  image: 1, M33 structure; 2, 125-kpc stream (stream A); 3, stream
  C; 4, eastern arc (stream D); 5, giant stellar stream; 6,
  northwest minor-axis stream; 7, southwest cloud. Features 1, 6
  and 7 (and part of 4) are new discoveries. Stellar sources were
  identified by using star-galaxy classification techniques
  described previously$^{14,16}$.  Candidate RGB stars were
  selected by their position in a colourÐmagnitude diagram
  relative to theoretical isochrones$^{29}$ for a 12-Gyr
  population at the distance of M31 (ref. 26), using only stellar
  sources with $i_0 < 23.5$. A projection of the stellar density
  distribution within a putative metallicity range $-2.5 < [Fe/H]
  < -0.6$\,dex was created with $0.02 \times 0.02$ pixels,
  smoothed with a Gaussian filter with dispersion 3 arcmin, and
  displayed with square-root scaling. The inset was created by
  combining, with a red-green-blue colour scheme, three such maps
  of M31 with metallicities of $-0.4 < [Fe/H] < +0.2$\,dex, $-1.3
  < [Fe/H] < -0.4$\,dex and $-2.3 < [Fe/H] < -1.3$\,dex,
  respectively (each with $0.01 \times 0.01$ degree pixels,
  smoothed with a Gaussian filter with 2 arcmin dispersion) and
  displayed with logarithmic scaling. Not all structures are
  visible with this (or any other) choice of metallicity cut,
  filter and scaling.}
\end{figure}

\begin{figure}
  \includegraphics[angle=0, width=14.cm]{figure2.ps}
\caption{{\bf Distribution of M31 dwarf galaxies. a},
    Points show the projected radial number density of dwarfs
    ($\Sigma$), derived for galaxies within 10\,degrees of M31 in the
    survey region. Horizontal error bars show the size of the bin,
    with points plotted at the mean galaxy radius in that
    bin. Vertical error bars are Poissonian. Galaxy names in black
    indicate the positions of the galaxies contributing to each bin;
    the remaining galaxies are named in grey. The radial ranges of the
    substructures indicated in Fig. 1 are also shown. {\bf b}, The
    luminosity distribution, using all galaxies within the
    survey. Error bars are Poissonian. The dashed line shows the
    best-fit Schechter function, with slope $\alpha = 0.98 \pm
    0.07$. Dark and light grey areas show 1$\sigma$ and 2$\sigma$
    deviations in $\alpha$, respectively. In both panels we are
    probably missing some satellites within $\sim 4$\,degrees of M31,
    where the high stellar density makes it difficult to detect faint
    satellites. We can detect galaxies brighter than $M_V < 26$,
    although derivation of the exact incompleteness levels remains to
    be calculated for the completed survey. To avoid these selection
    effects, the slope of the luminosity function is derived by using
    a $\chi^2$ fit of a Schechter function to galaxies brighter than
    $M_V < -8$, where incompleteness is not an issue, and we note that
    the extrapolation to lower luminosities fits these data
    well. Because of the lack of bright galaxies, we cannot
    independently derive $M^\star$. Instead, we fix $M^\star$ at a
    range of values and derive corresponding values of $\alpha$,
    finding that $\alpha$ is robust to within $1\sigma$ for $-19.5 >
    M^\star > -21.5$ (with a trend such that fainter values for
    $M^\star$ correspond to lower values for $\alpha$). The quoted
    value of $\alpha$ corresponds to $M^\star = -21$.}
\end{figure}

\begin{figure}
  \includegraphics[angle=270, width=12.cm]{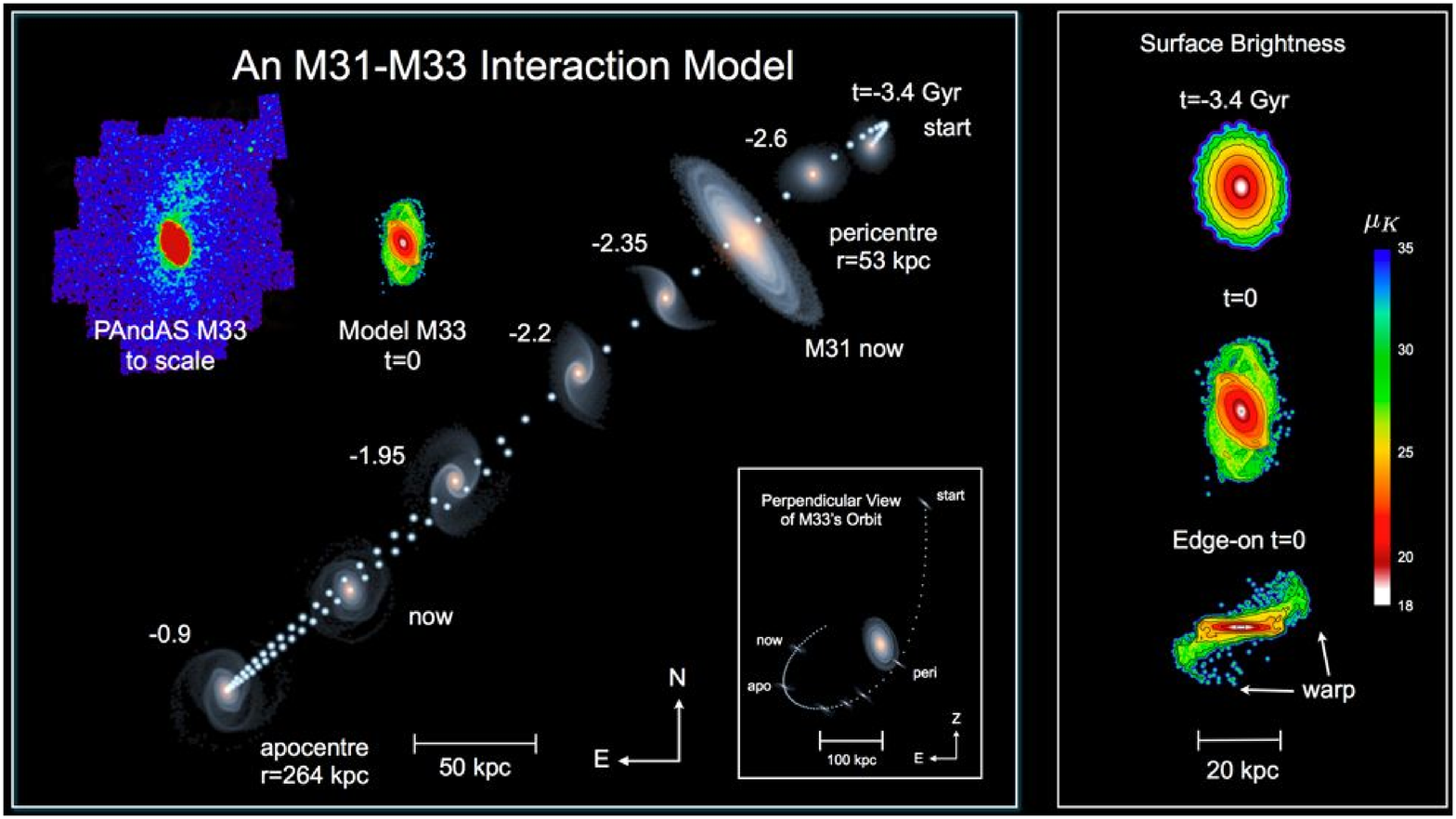}
\caption{{\bf An M31 - M33 interaction. a}, The evolution of
M33 about M31 along an orbit consistent with the angular positions,
distances$^{26}$ and radial velocities of M31 and M33 and with M33's
proper motion$^{23}$. Here, M33 starts 3.4\,Gyr ago at a distance of
$r < 200$\,kpc on the far side of M31 falling down the line of sight
to the Milky Way. After 800 Myr, M33 reaches pericentre and proceeds
across our line of sight towards the southeast, reaching apocentre
about 900 Myr ago before falling back towards M31 to its current
position. Dots tracing the orbit are separated by 49 Myr to give a
sense of the speed along the orbit. The lower inset shows a
perpendicular view of M33's orbit. {\bf b}, Quantification of the
expected K-band surface brightness of M33 at different times from
face-on and edge-on perspectives. The inner red/orange region with
$\mu_K < 22$ mag\,arcsec$^{-2}$ defines the size of the usual optical
disk of M33 seen in images. The initial equilibrium models$^{30}$ for
the two galaxies consist of a disk, bulge and dark-matter halo with
structural parameters that accurately reproduce the observed surface
brightness profiles and rotation curves. Because the mass profile of
M31 beyond 100 kpc is not well constrained by observations, we appeal
to cosmological arguments$^{24}$ that predict a mass of $2.5 \times
10^{12}$ solar masses within $r < 280$\,kpc. See the Supplementary
Information for more details.}
\end{figure}

\newpage

\section*{Supplementary Information}

\noindent{\it 1. Galaxy Models}

We construct realistic, self-consistent models for M31 and M33 for use
in our simulations. They consist of an exponential disk, a Sersic
bulge, and a spherical cuspy dark matter halo. From these we construct
a self-consistent distribution function and so generate equilibrium
N-body systems for simulations.  The key structural parameters for the
model used in the simulation are given in Supplementary Table 1.  The
masses of the three components, $M_h$, $M_d$, and $M_b$ are given in
units of $10^{10}\,M_\odot$.  M33 is naturally truncated by the tidal
field of M31 at a radius of $\sim$50~kpc at the initial distance of
230~kpc in the model so we quote the M33 halo mass within this tidal
radius.  Note that for an isolated galaxy with M33's rotation curve,
we would normally infer a virial radius $>100$~kpc and virial mass
about twice that quoted in the table below so it is important to
truncate the model in this context.  The M31 model extends to a virial
radius of $r_{virial}=280$ kpc and we quote the halo mass inside this
radius.  The halo scale radius, $r_s$, the disk scale length, $R_d$,
and the effective radius of the bulge, $R_e$, are given in
kiloparsecs.  The Sersic index of the bulge is $n_s$. We have chosen a
suitable mass-to-light ratio for the disk components to ensure that
  the disks are both intrinsically stable against the formation of a
strong bar.  To probe the evolution of the outer disks, we truncate
the M33 model at 10 scale-lengths (13 kpc) and the M31 model at 7
scale-lengths (40 kpc).  These radii are larger than the classical
size of the optical disks that extend to approximately 5
scale-lengths. 

\begin{table}
\begin{tabular}{|c|cc|cc|ccc|}
\hline
{} & \multicolumn{2}{|c|}{Halo} &
\multicolumn{2}{|c|}{Disk} & \multicolumn{3}{|c|}{Bulge} \\
& $M_h$ & $r_s$ & $M_d$ & $R_d$ & $M_b$ & $R_e$ & $n_s$\\
\hline
M33 &   8 & 12 & 0.26 & 1.3 & 0.055 & 0.84 & 2.5 \\
M31 & 247 & 22 & 5.8 & 5.8 & 3.3 & 1.5 & 1.0 \\
\hline
\end{tabular}
\caption{Equilibrium galaxy parameters}
\end{table}

\noindent{\it 2. Orbital Model}

We set our galaxy models up on a orbit that will lead to an
interaction while satisfying the observed constraints.  We define the
orbit relative to the M31 centre.  The assumed distances imply a
current separation of M31 and M33 of 206~kpc.  We move from
heliocentric coordinates to a frame with an origin at the Galactic
centre and correct for the effect of Galactic rotation using standard
values of the rotation velocity and motion with respect to the Local
Standard of Rest.  The proper motion of M31 is as yet unmeasured so
the transverse velocity components of M31 are unconstrained.  However,
if we assume values of the M31 proper motion then we can fully
determine the current relative position and velocity and so determine
the orbit.  Finally, the inclination and position angles and direction
of rotation of M31 and M33 determine the orientation of their
respective disks in space.  We assume that relative orientations of
the two galaxies remain essentially unchanged during an interaction
implying that the initial orientation of the two disks is the same as
it is at the current epoch in an orbital model.

We now try to find an orbit (or family of orbits) that satisfy the
observed constraints.  The main unknowns are the M31 transverse
velocity components so we allow these to vary as free parameters.  We
first search the orbital space by integrating the motion of a M33 test
particle in the gravitational potential of the M31 model defined
above.  Our goal is to find orbits with initial conditions that permit
a strong enough tidal encounter to excite the observed features in
M33.  We therefore search for orbits that have close encounters.  In
test particle integrations, the observed constraints appear to limit
the pericenter distance to a value no smaller than $r_p\approx 40$
kpc.  In an effort to excite the strongest response we therefore
confined ourselves to orbits with $r_p<50$~kpc which narrows the range
of possible orbits considerably.  Most orbits tend to be confined to a
plane nearly parallel to our line of sight and most of the
interactions occur with the past few billion years.  The model
presented in this paper is representative of this family.

Dynamical friction is a further complication.  Its effect was
significant and affected our predictions of initial conditions based
on the test particle integrations above.  We therefore incorporate
dynamical friction into our orbital search calibrating the parameters
in the dynamical friction formula using the N-body simulations.  In
this way, it was possible to predict initial conditions for the
simulations that would result in the M31 and M33 systems ending up in
the current positions and relative velocities within the errors of the
known constraints and assumed proper motion of M31.

\noindent{\it 3. N-body Model}

\begin{table}
\begin{tabular}{|c|c|c|c|}
\hline
{} & \multicolumn{1}{|c|}{Halo} &
\multicolumn{1}{|c|}{Disk} & \multicolumn{1}{|c|}{Bulge} \\
& $N_h$ & $N_d$ & $N_b$ \\
\hline
M33 & 2M & 1M & 200K \\
M31 & 2M & 1M & 500K\\
\hline
\end{tabular}
\caption{Numerical resolution}
\end{table}

We construct N-body models with a total of 6.7 million particles
divided up according to Supplementary Table 2.  This numerical
resolution was adequate to illustrate the main features of a tidal
encounter though future work will increase particle numbers to resolve
finer details of the interaction.  The simulation was carried out
using a parallelized tree-code.

\noindent{\it 4. Uniqueness of the model}

The observed constraints on the relative position and velocity of M33
along with the assumption of a close encounter lead to a range of
models similar to the one presented in this paper.  Another implicit
assumption in determination of this orbit is the gravitational
potential of M31 with parameters described above.  The outer mass
profile and virial mass of M31 might be different and this would
change the detailed character of the orbit and interaction.  However,
the inner mass profile of M31 is well-constrained so the strength of
the tidal encounter will not be significantly different in models
where the outer mass profile varies.  So while the model presented
here is not unique it demonstrates with reasonable accuracy the effect
of a recent close encounter on the evolution of M33 in a plausible
orbit.

The model demonstrates that such an encounter will excite tidal tails
in the outer disk of M33 shortly after the encounter as well as a
warp.  As the galaxy proceeds along its orbit, these features wind up
leaving a disturbed outer disk.  The extent of these features will
depend in detail on the initial truncation radius of the disk and
pericentric separation.  Also, the phase of disk extension and warp
will depend on the exact timing of the orbit.  Matching up all these
details simultaneously with all constraints is a difficult challenge,
but the model presented here demonstrates quite clearly that a
disturbance in the stellar disk and a warp with similar properties to
the observed features can be generated by a recent interaction that is
sufficiently close.

\newpage

\noindent1. White, S. D. M. \& Rees, M. J. Core condensation in heavy halosÑatwo-stage theory
for galaxy formation and clustering. Mon. Not. R. Astron. Soc. 183, 341-358 (1978).

\noindent2. Bullock, J. S. \& Johnston, K. V. Tracing galaxy formation with stellar halos. I.
Methods. Astrophys. J. 635, 931-949 (2005).

\noindent3. Abadi, M. G., Navarro, J. F. \& Steinmetz, M. Stars beyond galaxies: the origin of
extended luminous haloes around galaxies. Mon. Not. R. Astron. Soc. 365,
747-758 (2006).

\noindent4. Johnston, K. V. et al. Tracing galaxy formation with stellar halos. II. Relating
substructure in phase and abundance space to accretion histories. Astrophys. J.
689, 936-957 (2008).

\noindent5. Eggen, O. J., Lynden-Bell, D. \& Sandage, A. R. Evidence from the motions of old
stars that the Galaxy collapsed. Astrophys. J. 136, 748-766 (1962).

\noindent6. Searle, L. \& Zinn, R. Compositions of halo clusters and the formation of the
galactic halo. Astrophys. J. 225, 357-379 (1978).

\noindent7. Ibata, R. A., Gilmore, G. \& Irwin,M. J. A dwarf satellite galaxy in Sagittarius. Nature
370, 194-196 (1994).

\noindent8. Klypin, A., Kravtsov, A. V., Valenzuela, O. \& Prada, F. Where are the missing
galactic satellites? Astrophys. J. 522, 82-92 (1999).

\noindent9. Moore, B. et al. Dark matter substructure within galactic halos. Astrophys. J. 524,
L19-L22 (1999).

\noindent10. Bullock, J. S., Kravtsov, A. V. \& Weinberg, D. H. Reionization and the abundance of
galactic satellites. Astrophys. J. 539, 517-521 (2000).

\noindent11. Kravtsov, A. V., Gnedin, O. Y. \& Klypin, A. A. The tumultuous lives of galactic
dwarfs and the missing satellites problem. Astrophys. J. 609, 482-497 (2004).

\noindent12. Belokurov, V. et al. The field of streams: Sagittarius and its siblings. Astrophys. J.
642, L137-L140 (2006).

\noindent13. Ibata, R., Irwin, M., Lewis, G., Ferguson, A. M. N. \& Tanvir, N. A giant stream of
metal-rich stars in the halo of the galaxy M31. Nature 412, 49-52 (2001).

\noindent14. Ferguson, A. M. N., Irwin,M. J., Ibata, R. A., Lewis, G. F. \& Tanvir, N. R. Evidence for
stellar substructure in the halo and outer disk of M31. Astron. J. 124, 1452-1463
(2002).

\noindent15. Martin, N. F. et al. Discovery and analysis of three faint dwarf galaxies and a
globular cluster in the outer halo of the Andromeda galaxy. Mon. Not. R. Astron.
Soc. 371, 1983-1991 (2006).

\noindent16. Ibata, R. et al. The haunted halos of Andromeda and Triangulum: a panorama of
galaxy formation in action. Astrophys. J. 671, 1591-1623 (2007).

\noindent17. McConnachie, A. W. et al. A trio of new Local Group galaxies with extreme
properties. Astrophys. J. 688, 1009-1020 (2008).

\noindent18. Johnston, K. V., Hernquist, L. \& Bolte, M. Fossil signatures of ancient accretion
events in the halo. Astrophys. J. 465, 278-287 (1996).

\noindent19. Thilker, D. A. et al. On the continuing formation of the Andromeda galaxy:
detection of HI clouds in the M31 halo. Astrophys. J. 601, L39-L42 (2004).

\noindent20. Schechter, P. An analytic expression for the luminosity function for galaxies.
Astrophys. J. 203, 297-306 (1976).

\noindent21. Rogstad, D. H., Wright, M. C. H. \&  Lockhart, I. A. Aperture synthesis of neutral
hydrogen in the galaxy M33. Astrophys. J. 204, 703-711 (1976).

\noindent22. Corbelli, E. \& Schneider, S. E. A warped disk model for M33 and the 21 centimeter
line width in spiral galaxies. Astrophys. J. 479, 244-257 (1997).

\noindent23. Brunthaler, A., Reid, M. J., Falcke, H., Greenhill, L. J. \& Henkel, C. The Geometric
distance and proper motion of the Triangulum Galaxy (M33). Science 307,
1440-1443 (2005).

\noindent24. Loeb, A., Reid, M. J., Brunthaler, A. \& Falcke, H. Constraints on the proper motion
of the Andromeda galaxy based on the survival of its satellite M33. Astrophys. J.
633, 894-898 (2005).

\noindent25. Dubinski, J. A parallel tree code. N. Astron. 1, 133-147 (1996).

\noindent26. McConnachie, A. W. et al. Distances and metallicities for 17 Local Group galaxies.
Mon. Not. R. Astron. Soc. 356, 979-997 (2005).

\noindent27. Richardson, J. C. et al. The nature and origin of substructure in the outskirts of
M31. I. Surveying the stellar content with the Hubble Space Telescope Advanced
Camera for Surveys. Astron. J. 135, 1998-2012 (2008).

\noindent28. Ibata, R. et al. On the accretion origin of a vast extended stellar disk around the
Andromeda galaxy. Astrophys. J. 634, 287-313 (2005).

\noindent29. Dotter, A. et al. The Dartmouth stellar evolution database. Astrophys. J. Suppl. Ser.
178, 89-101 (2008).

\noindent30. Widrow, L. M., Pym, B. \& Dubinski, J. Dynamical blueprints for galaxies. Astrophys.
J. 679, 1239-1259 (2008).

\newpage

\begin{addendum}
 \item We thank the entire staff at the Canada - France - Hawaii Telescope for
taking the data, for initial processing with Elixir and for their
continuing support throughout this project. A.M.N.F. and A.D.M. are
supported by a Marie Curie Excellence Grant from the European
Commission under contract MCEXT-CT-2005-025869. G.F.L. thanks the
Australian Nuclear Science and Technology Organisation (ANSTO) for
supporting his involvement in PAndAS through its Access to Major
Research Facilities Program (AMRFP). R.M.R.  acknowledges grant
AST-0709479 from the National Science Foundation, and grants GO-9453,
GO-10265 and GO-10816 from the Space Telescope Science Institute. The
image of M33 overlaid in Fig. 1 is reproduced by courtesy of
T. A. Rector and M. Hanna.
 \item[Competing Interests] The authors  have no
competing financial interests.
\item[Author contributions] All authors assisted in the
development and writing of the paper. In addition, A.W.M. is the
Principal Investigator of PAndAS; M.J.I. led the data processing
effort; R.A.I. was the Principal Investigator of an earlier CFHT
MegaPrime/MegaCam survey, which PAndAS builds on (which included
S.C.C., A.M.N.F., M.J.I., G.F.L., N.F.M., A.W.M. and N.T.); J.D.,
L.M.W. modelled the M31 - M33 interaction; N.F.M. had a lead role in the
study of the dwarf galaxies; P.C.  assisted with constructing the
luminosity function; and A.L.D. developed the theoretical isochrones.
 \item[Correspondence] Reprints and permissions information
is available at www.nature.com/reprints. Correspondence and requests
for materials should be addressed to
A.W.M. (alan.mcconnachie@nrc-cnrc.gc.ca).
\end{addendum}


\end{document}